\pdfoutput=1
\documentclass[usenatbib]{mn2e}
\usepackage{apjfonts,amssymb,amsmath,ctable,url}
\usepackage[implicit=false,breaklinks,colorlinks,citecolor=blue]{hyperref}

\newcommand{\pdot}{\dot{\bf p}}
\newcommand{\pdota}{(\pdot_{\nu})_{a}}
\newcommand{\pdotab}{(\pdot_{\nu})_{ab}}
\newcommand{\pdotba}{(\pdot_{\nu})_{ba}}
\newcommand{\mfp}{\lambda_{\rm MFP}}
\newcommand{\Va}{\Delta V_{a}}

\newcommand{\gizmourl}{\href{http://www.tapir.caltech.edu/~phopkins/Site/GIZMO.html}{\url{http://www.tapir.caltech.edu/~phopkins/Site/GIZMO.html}}}

\newcommand{\msun}{M_{\sun}}

\newcommand{\apjl}{ApJL}
\newcommand{\apjs}{ApJS}

\newcommand{\aap}{A\&A}

\newcommand\plotonesize[2]
 {\centering \leavevmode \includegraphics[width={#2\columnwidth}]{#1}}
\newcommand{\plotsidesize}[2]
 {\centering \leavevmode \includegraphics[width={#2\textwidth}]{#1}}
\newcommand{\acknowledgments}{\begin{small}\section*{Acknowledgments}\end{small}}
\newcommand\altaffilmark[1]{$^{#1}$}
\newcommand\altaffiltext[1]{$^{#1}$}
\title[Photon Momentum Coupling]{Numerical Problems in Coupling Photon Momentum (Radiation Pressure) to Gas
\vspace{-0.5cm}}

\vspace{-0.2cm}
\author[Hopkins \&\ Grudi\'{c}]{
\parbox[t]{\textwidth}{ 
Philip F.~Hopkins\thanks{E-mail:phopkins@caltech.edu}\altaffilmark{1},
Michael Y.~Grudi\'{c}\altaffilmark{1}
} 
\vspace*{6pt} \\
\altaffiltext{1}{TAPIR, Mailcode 350-17, California Institute of Technology, Pasadena, CA 91125, USA} %\\
\vspace{-0.5cm}
}

\date{Submitted to MNRAS, February 2018\vspace{-0.6cm}}
\begin{document}
\maketitle
\label{firstpage}

\vspace{-0.2cm}
\begin{abstract}
%\vspace{-0.2cm}
Radiation pressure (RP; or photon momentum absorbed by gas) is important in a tremendous range of astrophysical systems. But we show the usual method for assigning absorbed photon momentum to gas in numerical radiation-hydrodynamics simulations (integrating over cell volumes or evaluating at cell centers) can severely under-estimate the RP force in the immediate vicinity around un-resolved (point/discrete) sources (and subsequently under-estimate its effects on bulk gas properties), unless photon mean-free-paths are highly-resolved in the {\em fluid} grid. The existence of this error is {\em independent} of the numerical radiation transfer (RT) method (even in exact ray-tracing/Monte-Carlo methods), because it depends on how the RT solution is interpolated back onto fluid elements. Brute-force convergence (resolving mean-free paths) is impossible in many cases (especially where UV/ionizing photons are involved). Instead, we show a ``face-integrated'' method -- integrating and applying the momentum fluxes at interfaces between fluid elements -- better approximates the correct solution at all resolution levels. The ``fix'' is simple and we provide example implementations for ray-tracing, Monte-Carlo, and moments RT methods in both grid and mesh-free fluid schemes. We consider an example of star formation in a molecular cloud with UV/ionizing RP. At state-of-the-art resolution, cell-integrated methods under-estimate the net effects of RP by an order of magnitude, leading (incorrectly) to the conclusion that RP is unimportant, while face-integrated methods predict strong self-regulation of star formation and cloud destruction via RP.
\end{abstract}

\begin{keywords}
methods: numerical --- radiation: dynamics --- hydrodynamics --- galaxies: formation --- stars: formation --- galaxies: active\vspace{-0.5cm}
\end{keywords}

\section{Introduction}
\label{sec:intro}

Radiation-hydrodynamics (RHD) is fundamental to the behavior of a huge range of astrophysical systems, including galaxies; star-forming giant molecular clouds (GMCs) and star clusters; the inter-stellar (ISM) circum-galactic (CGM) and inter-galactic (IGM) medium; proto-planetary disks (PPDs) and sites of planet formation; stellar structure and evolution; accretion physics, compact objects, and active galactic nuclei (AGN); supernovae; and more. In most of these systems the ``radiation pressure forces'' -- meaning both optically thin transfer of momentum from the radiation field to the gas via absorbed photon momentum, and optically thick (multiple scattering) local momentum transfer generating a true ``pressure'' -- cannot be neglected. However, because these are complicated, non-linear, turbulent, multi-physics systems, numerical simulations are often necessary, and RHD is a computationally challenging problem for many reasons. As such, a great deal of literature has focused on numerical methods for solving (or approximately solving) the radiative transfer equation (RTE) on-the-fly in simulations \citep[for some recent examples, see e.g.][]{hopkins:rad.pressure.sf.fb,davis:2012.rhd.short.characteristics,kuiper:2012.rad.pressure.outflow.vs.rt.method,bate:2012.rmhd.sims,wise:2012.rad.pressure.effects,pluto:2013.fld.implicit,rosdahl:2013.m1.ramses,davis:2014.rad.pressure.outflows,gonzalez:2015.fld.rhd,roth:2015.implicit.mc,tominaga:2015.spherical.harmonic.rhd,buntemeyer:2016.art,zhang:2017.rhd.dusty.winds,rosen:2017.harm.rhd,kim:2017.art.uv.starclusters,foucart:2018.pic.closure}

But comparably little attention has been paid to how, {\em given} a solution to the RTE, radiation pressure forces are coupled {\em onto the gas or fluid} \citep[although see e.g.][]{lowrie:1999.radiation.hydro.coupling}. In this paper, we show that the most common methods used in the literature can artificially suppress the radiation pressure forces in gas elements nearby un-resolved (discrete or point or ``sub-grid'') sources by orders-of-magnitude, unless the photon mean free paths are very well-resolved. This is especially problematic for physical systems with many discrete sources (e.g.\ star and galaxy formation), or where the source emission region is optically-thick and poorly-resolved (e.g.\ AGN and black hole emission), and in those regimes can be particularly important for the large-scale, bulk effects of radiation on the system. But in many of these contexts, the mean-free-paths of especially ionizing and UV photons in the dense gas around sources (where radiation pressure is most important) can be many orders-of-magnitude smaller than state-of-the-art numerical resolution. 

In \S~\ref{sec:physics} we briefly review the equations solved for the RTE and radiation pressure forces, and common classes of numerical implementations. \S~\ref{sec:single.scattering} considers the coupling of photon momentum in the single-scattering limit, and shows how the typical ``cell-integrated'' approach fails (\S~\ref{sec:single.scattering:cell.integrated}), and why the resolution requirements to ``brute force'' convergence with this method are impractical (\S~\ref{sec:resolution}). We propose instead, in \S~\ref{sec:face}, an alternative, face-integrated approach, which resolves these errors and properly treats the absorbed photon momentum around sources even in the limit where photon mean free paths are unresolved. We show how to implement this in various (ray-tracing, Monte-Carlo, moments-based) RTE methods, and various (fixed grid, mesh-free) hydrodynamic methods (additional details in App.~\ref{sec:appendix.face.construction}). \S~\ref{sec:multiple.scattering} generalizes to multiple-scattering, and \S~\ref{sec:test.problem} demonstrates how these errors can have a dramatic impact on real astrophysical problems, using a simulation of star formation in a molecular cloud as an example. We conclude in \S~\ref{sec:discussion}.

\vspace{-0.5cm}
\section{Review of RHD Methods}
\label{sec:physics}

Recall, the radiative transfer equation (RTE), to the order of interest (neglecting higher-order relativistic \&\ Hubble-flow effects unimportant here), is 
\begin{align}
\label{eqn:RTE} \left(\frac{1}{c}\,\frac{\partial }{\partial t} + \hat{\Omega}\cdot\nabla\right)\,I_{\nu}  &= 
j_{\nu} - \alpha_{\nu}\,I_{\nu} \\
\nonumber & - \int d\Omega^{\prime}\,[ I_{\nu}\,\sigma_{\nu}(\Omega,\,\Omega^{\prime}) - I_{\nu}(\Omega^{\prime})\,\sigma_{\nu}(\Omega^{\prime},\,\Omega)]
\end{align}
\citep{mihalas:1984oup..book.....M}, where $c$ is the speed of light, $\hat{\Omega}$ the unit vector in the direction of some differential solid angle $d\Omega$, $\alpha_{\nu}$ and $\sigma_{\nu}$ the absorption and scattering coefficients, $j_{\nu}$ the emissivity, and $I_{\nu}$ the intensity. Ignoring scattering and sources, this is
\begin{align}
\left(\frac{1}{c}\,\frac{\partial }{\partial t} + \hat{\Omega}\cdot\nabla\right)\,I_{\nu} &= -\rho\,\kappa_{\nu}\,I_{\nu} = - \frac{I_{\nu}}{\mfp} 
\end{align}
where $\rho$ is the gas density, $\kappa_{\nu}$ the opacity, and $\mfp \equiv 1/(\rho\,\kappa_{\nu})$ the photon mean free path (MFP). Within an infinitesimally small differential volume $d{\rm Vol} = d^{3}{\bf x}$, the rate of momentum transfer from radiation to gas is: 
\begin{align}
\label{eqn:pdoc.diff} \frac{d\, \pdot}{d^{3}{\bf x}} &= \frac{\partial (\rho\,{\bf u})}{\partial t} =  \int d \nu \, d\Omega \,\frac{\rho\,\kappa_{\nu}\,I_{\nu}\,\hat{\Omega}\,d{\Omega}}{c} = \int d\nu\, \frac{\rho\,{\kappa_{\nu}}\,{\bf F}_{\nu}}{c} 
\end{align}
where ${\bf F}_{\nu} \equiv \int d\Omega\,I_{\nu}\hat{\Omega}$ is the flux. We can also define the energy density $e_{\nu} \equiv c^{-1}\,\int d\Omega\,I_{\nu}$, and source function $S_{\nu} \equiv \int d\Omega\,j_{\nu}$.

\begin{figure*}
    \plotsidesize{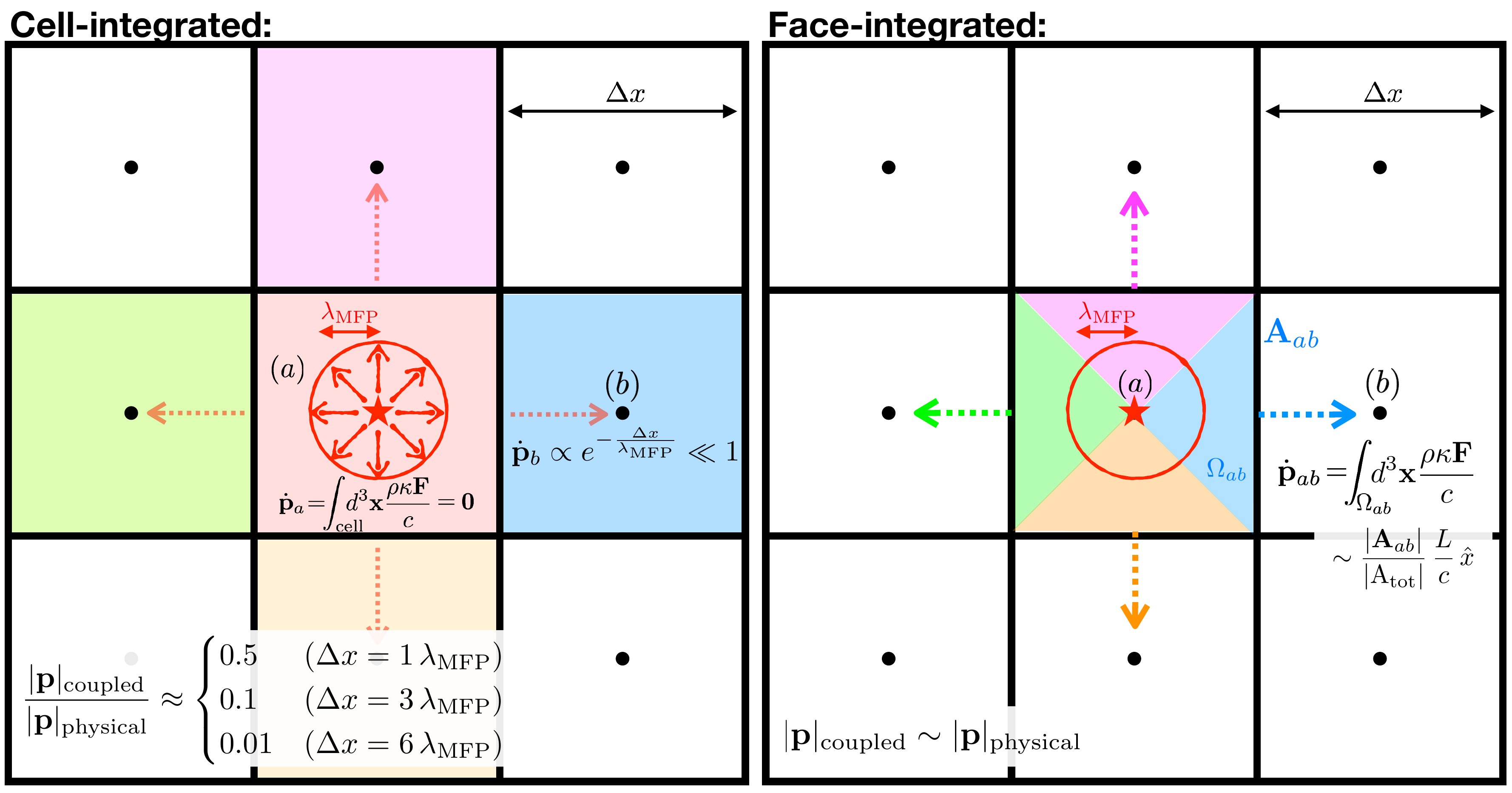}{0.99}
    \vspace{-0.25cm}
    \caption{Cartoon illustrating the key errors discussed here (and their solution). A single monochromatic radiation source (red star) sits at the center (cell $a$) of the fluid grid (cells with side-length $\Delta x$), emitting an isotropic flux ${\bf F}=\hat{r}\,L/(4\pi\,r^{2})$ of photons which are singly-scattered in a medium with constant density $\rho$ and opacity $\kappa$ (photon mean-free-path $\mfp=1/\rho\,\kappa$). We assume the RTE (Eq.~\ref{eqn:RTE}) is solved perfectly (${\bf F}$ is known exactly everywhere). 
    {\em Left:} Cell-integrated coupling. A momentum flux $\dot{\bf p}_{a,\,b}$ is assigned to cell $a$ or $b$ by integrating the absorbed photon momentum density $=\rho\,\kappa\,{\bf F}/c$ over all points in the volume/domain of $a$ or $b$ (Eq.~\ref{eqn:cell.integrated.dp}). If $\Delta x \gg \mfp$, almost all photons are absorbed in cell $a$, but since these are emitted (by definition) isotropically, the {\em net} $\dot{\bf p}_{a} = {\bf 0}$. Some coherent momentum is imparted to cell $b$ but it is lower by a factor $\sim \exp{(-\Delta x/\mfp)} \ll 1$. As a result, the radiation pressure will do nothing to the gas, even in the limit $L\rightarrow\infty$. 
    {\em Right:} Face-integrated coupling. Instead of averaging over volume, we integrate the absorbed photon momentum density in the domain $\Omega_{ab}$ flowing towards/through each face ${\bf A}_{ab}$ between neighboring fluid elements $a$ and $b$ (Eq.~\ref{eqn:face.integrated.dp}), and assign this as a flux or force between elements (at the face). Even when $\Delta x \gg \mfp$, the correct momentum ``outward'' towards each face from the source is now recovered. 
    \label{fig:single.scattering.demo}}
\end{figure*}

A range of methods in the literature attempt to solve Eq.~\ref{eqn:RTE} ``on the fly'' in simulations, subject to various approximations. The problems highlighted here apply to {\em all} methods, in principle, but because details of the necessary correction differ, we briefly review broad classes of popular methods for solving the RTE.

\vspace{-0.5cm}
\subsection{Collisionless (Ray-Tracing \&\ Monte-Carlo) Methods}

Ray-tracing with ``long characteristics'' (RT-LC) and/or Monte Carlo (MC) methods, which explicitly track $I_{\nu}$ along discrete ``rays'' or photon packets from all sources, can (at least in principle) exactly solve the RTE. This is often prohibitively expensive so it is common to make a variety of approximations, e.g.\ (1) using a limited (finite) number of rays/photon packets; (2) assuming either an infinite-speed-of-light/equilibrium solution (dropping the $c^{-1}\,\partial I_{\nu}/\partial t$ term) or ``reduced speed of light'' ($c \rightarrow \tilde{c} \ll c$) in explicit non-equilbrium solutions, both of which allow for larger timesteps; (3) neglecting scattering (or assuming it is isotropic); (4) neglecting velocity-dependence in $\kappa_{\nu}$ (for continuum transfer) and/or higher-order relativistic terms; (5) limiting to a small number of discrete frequency ``bins'' by integrating over some finite $\Delta \nu$; (6) replacing direct rays from all sources in the volume to all points with ``adaptive ray tracing'' (ART; \citealt{abel:2002.adaptive.ray.tracing}) where rays are ``split'' or ``merged'' to sample the domain more/less accurately in some regions, or ``short-characteristics'' (RT-SC; \citealt{olson:1987.short.char.rhd}) where only a fixed set of angles within each cell (which must be interpolated between cells) are used; (7) following direct rays but collapsing non-local shadowing/extinction to a local isotropic region around each source (LEBRON; \citealt{hopkins:2013.fire,hopkins:fire2.methods}); (8) treating absorption and re-emission as elastic scattering (in e.g.\ ``implicit Monte Carlo'' methods; \citealt{fleck:1971.implicit.monte.carlo}). 

For our purposes, these approximations are not important -- these methods are all similar in the key respects that they explicitly follow $I_{\nu}$ along different angles $\hat{\Omega}$, and (at least in principle) $I_{\nu}$ can have structure below the fluid grid scale.

\begin{figure*}
    \plotsidesize{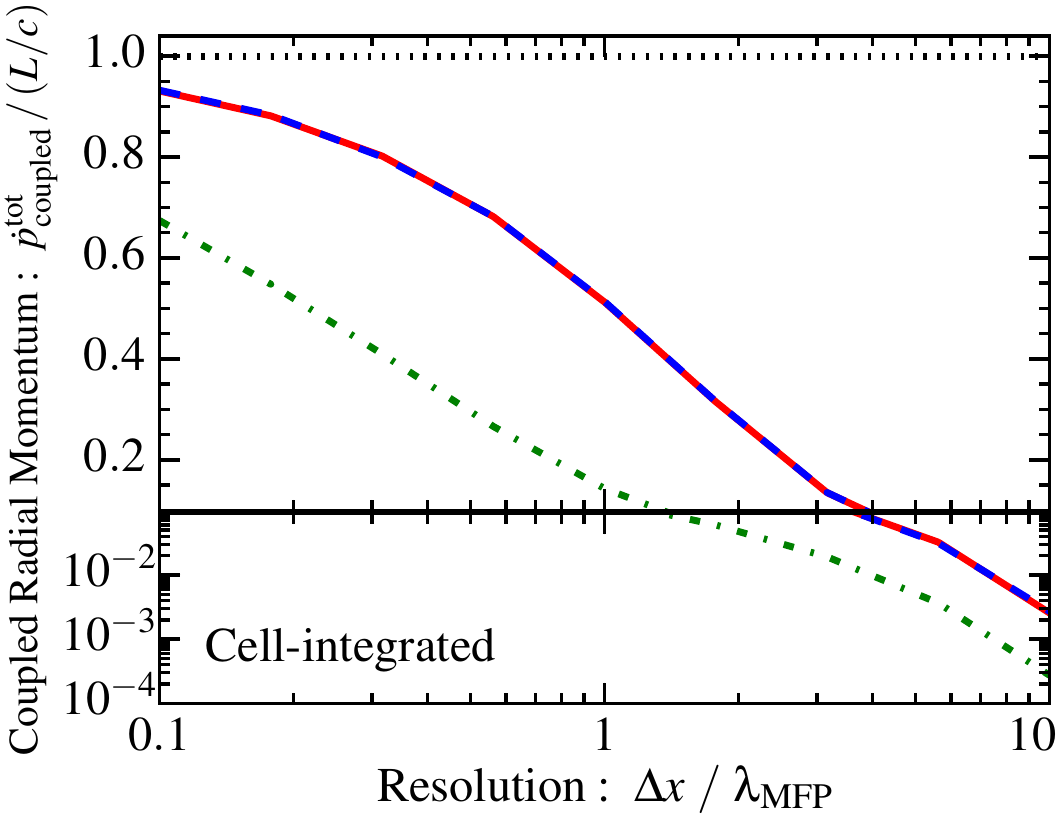}{0.50}
    \plotsidesize{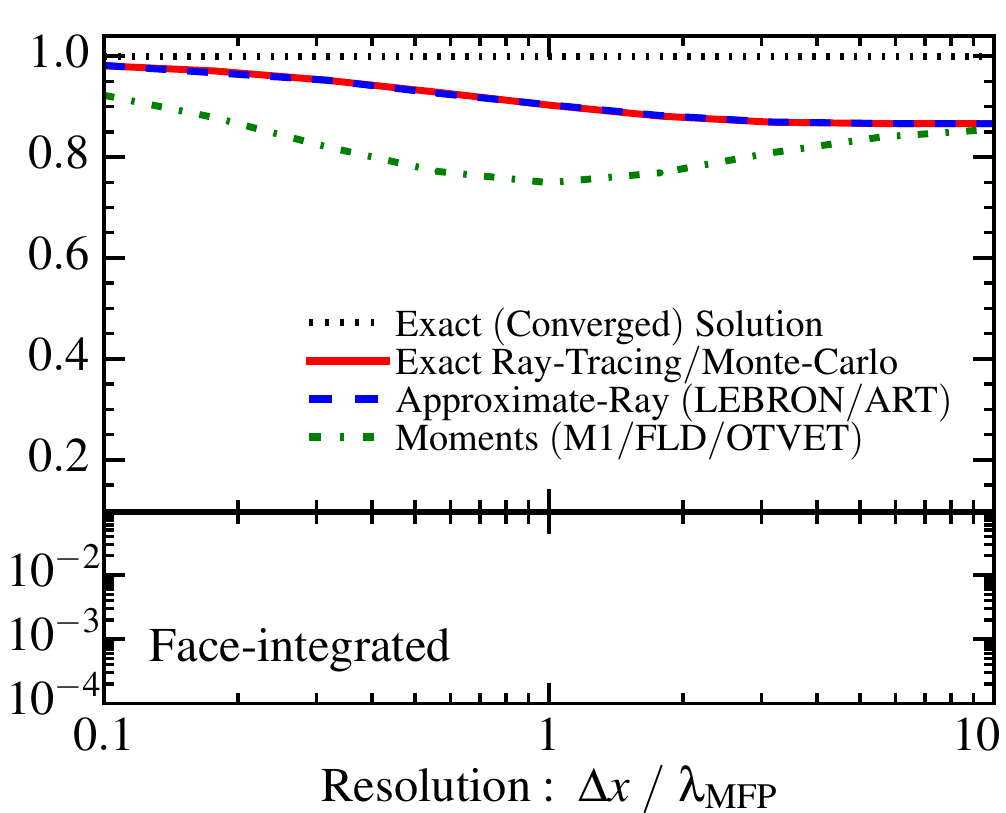}{0.477}
    \vspace{-0.25cm}
    \caption{Quantitative demonstration of the error from the simple test in Fig.~\ref{fig:single.scattering.demo}. We take that setup, place the single source at a random position on the regular Cartesian grid (averaging over $\sim1000$ positions), and solve the RTE (waiting for it to come into its equilibrium solution, with the gas properties fixed) using one of three methods: (1) an ``exact'' (e.g.\ infinite-resolution ray-tracing or Monte-Carlo) method, where the flux is exact at all points; (2) an approximate ray-tracing method (using the ray-tracing in LEBRON or ART gives nearly-identical results here, both to each other and to exact RT methods); (3) a ``moments'' method (M1 here, but FLD and OTVET give identical results). We then calculate the total radial momentum flux, $\sum_{a}\,\dot{\bf p}_{a}$ summed over all cells (or faces) according to that RTE solution and either the cell-integrated (Eq.~\ref{eqn:cell.integrated.dp}; {\em left}) or face-integrated (Eq.~\ref{eqn:face.integrated.dp}; {\em right}) coupling methods. With cell-integrated methods the true RP force is suppressed by orders-of-magnitude, even given a perfect RT solution, if $\Delta x \gtrsim \mfp$ (in moments schemes, additional errors require $\sim 10\times$ higher resolution; see \S~\ref{sec:moments.additional.problems}). Face-integrated methods  give robust answers at all resolution levels (all solutions asymptote to $\approx 85\%$ of the full radial momentum as $\Delta x\rightarrow \infty$, but this is the correct geometric effect from the finite number of faces in a grid cell). 
    \label{fig:acc.calc}}
\end{figure*}

\vspace{-0.5cm}
\subsection{``Moments'' or ``Fluid'' Methods}
\label{sec:moments.overview}

In moments methods, one integrates the RTE over $\int \hat{\Omega}^{n}\,d\Omega$, with $n=0,\,1,\,...$, to take the $n^{\rm th}$ moments. The hierarchy of moment equations never closes (each depends on the next-higher moment), so one truncates the series by adopting an ad-hoc closure {\em ansatz}.

Flux-limited diffusion (FLD; \citealt{levermore:1981.fld}) closes the hierarchy at $n=0$, giving $\partial e_{\nu}/\partial t + \nabla \cdot {\bf F}_{\nu} = S_{\nu} - \rho\,\kappa_{\nu}\,\tilde{c}\,e_{\nu}$, with the {\em ansatz} ${\bf F}_{\nu} \rightarrow (\lambda_{L}\,\tilde{c}/\kappa_{\nu}\,\rho)\,\nabla\cdot (e_{\nu}\,\mathbb{D})$ with Eddington tensor $\mathbb{D}=\mathbb{I}/3$ ($\mathbb{I}$ is the identity matrix) and the ``flux-limiter'' $\lambda_{L} = 3\,(2+\mathcal{F})/(6+3\,\mathcal{F} + \mathcal{F}^{2})$ (with $\mathcal{F} \equiv |\nabla e_{\nu}|/(e_{\nu}\,\kappa_{\nu}\rho)$) chosen to interpolate between free-streaming and isotropic-diffusion-like behavior based on an estimate of the local optical depth. The ``optically thin variable Eddington tensor'' (OTVET; \citealt{gnedin.abel.2001:otvet}) approach is identical to FLD but with $\mathbb{D} = \hat{\bf F}_{\nu}^{\rm thin}\otimes \hat{\bf F}_{\nu}^{\rm thin}$ where $\hat{\bf F}_{\nu}^{\rm thin}$ is the flux that would be present with no obscuration anywhere. The ``moment-one'' (M1; \citealt{levermore:1984.FLD.M1}) method closes at $n=1$, keeping the energy equation above and adding the flux equation $\tilde{c}^{-1}\,\partial {\bf F}_{\nu}/\partial t  + \tilde{c}\,\nabla \cdot \mathbb{P}_{\nu} = -\rho\,\kappa_{\nu}\,{\bf F}_{\nu}$ (so ${\bf F}_{\nu}$ is explicitly evolved), with the {\em ansatz} $\mathbb{P}_{\nu} \rightarrow e_{\nu}\,\mathbb{D}_{\nu}$, $\mathbb{D}_{\nu} = (1-\chi_{\nu})\,\mathbb{I}/2 + (3\,\chi_{\nu}-1)\,(\hat{\bf F}_{\nu}\otimes \hat{\bf F}_{\nu})/2$, $\chi_{\nu} = (3+4\,f_{\nu}^{2}) / (5 + 2\sqrt{4-3\,f_{\nu}^{2}})$, $f_{\nu} \equiv |{\bf F}_{\nu}| / (e_{\nu}\,\tilde{c})$ which again interpolates between free-streaming and isotropic diffusion. Other assumptions for the closures are of course possible but change nothing for our study here.

Independent of the closure, the salient feature of these methods is that each moment is {\em single-valued} at a given ${\bf x}$ -- one has made the fluid approximation for the radiation. This gives the well-known result that moments methods do not, in fact, converge to the correct RTE solutions (e.g.\ anti-parallel photon streams will ``collide'') except in the limit when the optical depths are everywhere infinite. These are popular methods because of their simplicity, but introduce unique complications discussed below. 

%The most important feature of these moments methods, {\em independent} of the closure, is that by integrating over $\Omega$ each moment is {\em single-valued} at a given location ${\bf x}$ -- one has essentially therefore made the fluid approximation for the radiation. This has a number of consequences: for example, it is fundamentally impossible to capture the collisionless nature of photons, so e.g.\ two photon streams moving in opposite directions will (incorrectly) collide, ``shock,'' merge, and then diffuse out from the collision, rather than simply passing through one another. As a consequence, these methods can {\em never} converge to the correct solution for the RTE, except in the limit of infinitely high optical depth. We discuss them, however, because they are a widely-employed approximation, and their fluid nature requires additional corrections and care if one wishes to treat photon momentum accurately.

\vspace{-0.5cm}
\section{Errors \&\ Corrections: Single-Scattering}
\label{sec:single.scattering}

First consider ``single-scattering'': photons free-stream until they are absorbed (no re-emission or scattering). To demonstrate the numerical issues, consider an idealized problem shown in Fig.~\ref{fig:single.scattering.demo}. A single monochromatic isotropic source with luminosity $L=L_{\nu}$ sits at position ${\bf x}_{0}$, centered on a computational domain which we take to be a regular Cartesian grid with cells of side-length $\Delta x$ (our conclusions are independent of the mesh geometry but this is convenient). Furthermore, assume that (regardless of RT method used) the RTE is solved {\em perfectly} at all points ${\bf x}$ in the domain.

\vspace{-0.5cm}
\subsection{The Problem: Cell-Integrated Coupling}
\label{sec:single.scattering:cell.integrated}

In most RHD implementations, {\em regardless} of the method used to solve the RTE, the radiation force on the gas is computed by integrating the momentum of absorbed photons over the volume of a given domain (cell) ``$a$'':\footnote{Even simpler than the ``cell-integrated'' Eq.~\ref{eqn:cell.integrated.dp}, some RHD implementations (in e.g.\ moments methods or some SPH/finite-point methods) adopt a ``cell-centered'' approach where the acceleration ${\bf a}_{\nu} \equiv c^{-1} \rho\,\kappa_{\nu}\,{\bf F}_{\nu}$ is evaluated at the cell center/particle location ${\bf x}_{a}$ and then assigned to the whole cell/particle. As this is equivalent to evaluating Eq.~\ref{eqn:cell.integrated.dp} using only the cell-centered ${\bf F}_{\nu}\rightarrow {\bf F}_{\nu}({\bf x}_{a})$ (likewise for $\rho$, $\kappa$), it produces the same errors as cell-integrated approaches. In addition it is noisier than cell or face-integrated methods, and can violate linear momentum conservation, so we will not discuss this particular case in more detail.}
\begin{align}
\label{eqn:cell.integrated.dp} \pdota &= \int_{\Va} d^{3}{\bf x}\,\frac{\rho\, \kappa_{\nu} {\bf F}_{\nu} }{c}
\end{align}
so the total momentum $\Delta {\bf p}_{a} = \int_{\Delta\nu}\int_{\Delta t} d\nu\,dt\,\pdota$ is simply added to the gas in $a$ each timestep $\Delta t$. For discrete (e.g.\ Monte Carlo) methods, this integral is given by summing over all absorptions in a given cell: $\Delta {\bf p}_{a} = c^{-1}\,\sum_{j}\,E_{\gamma}^{j,\,{\rm abs}}\,\hat{\Omega}_{j}$, where $E_{\gamma}^{j,\,{\rm abs}}$ is the total photon energy absorbed in photon packet/ray $j$, within domain $a$, over timestep $\Delta t$.

Now consider the case where the photon MFP is un-resolved by the {\em fluid} grid, i.e.\ $\Delta x \gg \mfp$. Essentially all the photons should therefore be absorbed inside the single cell $a$ surrounding the source. Eq.~\ref{eqn:cell.integrated.dp} then trivially evaluates to $\pdota = \Delta {\bf p}_{a} = {\bf 0}$ because the source emits isotropically. 

In the absence of other physics, the system would remain (incorrectly) perfectly static as $t\rightarrow \infty$. The correct solution (with negligible gravity and pressure forces) is that a shell of material moves away from the source, sweeping up gas (leaving an empty cavity) with radial momentum flux $= L/c$ (so the total {\em radial} momentum $\int d^{3}{\bf x}\,\rho\,{\bf v}\cdot\hat{r} = L\,t/c$). If $L$ and $\rho$ are constant the shell should expand indefinitely with radius $r(t) = (3\,L\,t^{2}/2\pi\,c\,\rho)^{1/4}$.

\vspace{-0.5cm}
\subsection{Solution 1: Increasing Resolution (Is Not Practical)}
\label{sec:resolution}

One solution to the problem above is to increase the resolution of the {\em hydrodynamic} grid until the photon mean-free paths are all well-resolved, i.e.\ $\Delta x \ll \mfp$.\footnote{Another possibility would be to artificially decrease the opacity in different bands, so that mean free paths are always resolved. This amounts to capping the opacity at $\kappa_{\nu}^{\rm cap} \ll 1/(\rho\,\Delta x)$ (in some cases accomplished by treating cells around sources as $\kappa_{\nu}=0$ ``ghost zones''). This immediately creates a number of problems: (1) Most important, this can lead to photons being absorbed at the wrong physical locations, far further from their sources than they should be. (2) It is very difficult to implement such a prescription in anything but an idealized simulation, without risking allowing photons to escape entirely from dense regions where they should have been absorbed (e.g.\ ``under-shooting'' the opacity). (3) The central cell around the source is still not correctly being ``swept up'' in a shell in the problem described above, because the photons are free-streaming out to neighboring cells. So a shell will form, but only external to a central dense region from which the photons escape -- thus the solution at finite resolution will still not correctly represent the converged solution.} This would eventually converge to the correct behavior (assuming a perfect RTE solution). 

But this is not possible for many real simulations. For example, in simulations of star formation, galaxies, or AGN (outside the accretion disk), the single-scattering absorption comes primarily from neutral hydrogen absorption of ionizing photons, with $\kappa_{\nu} \sim 4\times10^{6}\,{\rm cm^{2}\,g^{-1}}$ (for neutral gas), and/or near-UV with $\kappa_{\nu} \gtrsim 3000\,{\rm cm^{2}\,g^{-1}}$, in relatively dense gas around the sources (e.g.\ $n \sim \rho/m_{p} \gtrsim 10^{2}-10^{4}\,{\rm cm^{-3}}$, in HII regions, or $\sim 10^{6}-10^{12}\,{\rm cm^{-3}}$ in the obscuring ``torii'' around AGN). In a uniform grid, this would require 
\begin{align}
\label{eqn:spatial.res} \Delta x \ll \mfp \sim 5\times10^{-6}\,n_{4}^{-1}\,{\rm pc}
\end{align}
(where $n_{4}=n/10^{4}\,{\rm cm^{-3}}$) for ionizing photons in neutral gas. Even for just near-UV photons this gives $\Delta x \ll  0.006\,n_{4}^{-1}\,{\rm pc}$, so a grid of $\sim 3100^{3}\sim 10^{10.5}$ elements is needed for a $\sim 10\,$pc-radius GMC (even if the maximum density were capped at $\sim 10^{4}\,{\rm cm^{-3}}$). In Lagrangian or AMR methods where the mass resolution $\Delta m$ is approximately fixed and spatial resolution is adaptive ($\Delta x \approx (\Delta m/\rho)^{1/3}$), the required mass resolution would be 
\begin{align}
\label{eqn:mass.res} \Delta m \ll \mfp^{2}/\kappa_{\nu} \sim 3\times10^{-14}\,n_{4}^{-2}\,\msun 
\end{align}
(for ionizing photons in neutral gas) or $\Delta m \ll 6\times10^{-5}\,n_{4}^{-2}\,\msun$ (for near-UV). This is wildly beyond the resolution of state-of-the-art simulations. 

If we consider absorption of hard radiation in the vicinity of accreting supermassive black holes (where the largest spatial scales of interest are similar, but the densities and opacities are even higher), the resolution-discrepancy only becomes more severe.

\vspace{-0.5cm}
\subsection{Solution 2: Face-Integrated Coupling}
\label{sec:face}

%\subsubsection{Method Overview \&\ Momentum-Coupling}

Consider instead a {\em face-integrated} momentum coupling. Instead of considering only the volume element $a$, note that $a$ is surrounded by a set of faces\footnote{In ``mesh-free'' methods (e.g.\ SPH, FPM) one can always define ``effective'' faces by reference to equation-of-motion and point locations: we provide a generic method for this in Appendix~\ref{sec:appendix.face.construction}.} $ab$ (between domain $a$ and each neighboring element $b$) each with an oriented vector area ${\bf A}_{ab}$. Now simply integrate the absorbed flux moving ``towards'' each face ${\bf A}_{ab}$: 
\begin{align}
\label{eqn:face.integrated.dp}  \pdotab &= \int_{\Va} d^{3}{\bf x}\,\frac{\rho\,\kappa_{\nu} {\bf F}_{\nu}}{c}\,\Theta_{ab}({\bf x},\,\hat{\bf F}_{\nu}) 
 = \int_{\Va,\,\Omega_{ab}} d^{3}{\bf x}\,\frac{\rho\,\kappa_{\nu} {\bf F}_{\nu}}{c}
\end{align}
where $\Theta_{ab}({\bf x},\,\hat{\bf F}_{\nu}) = 1$ if the face ${\bf A}_{ab}$ is the ``intercepted face'' (i.e.\ if ${\bf A}_{ab}$ is the first face crossed, along the ray originating at ${\bf x}$ in direction $\hat{\bf F}_{\nu}$), and $=0$ otherwise. Equivalently, $\Theta=1$ if $\hat{\bf F}_{\nu}$ ``points to'' face $ab$ from ${\bf x}$ within $a$. Thus this is the integral over the domain $\Va[\Omega_{ab}]$ where $\Omega_{ab}$ is the range of solid angle subtended by face ${\bf A}_{ab}$ from ${\bf x}$ (i.e.\ angles where $\Theta({\bf x},\,\hat{\Omega})=1$). In (exact or approximate) ray-tracing methods where absorption is calculated along rays with fixed $\hat{\Omega}$, it is straightforward to evaluate Eq.~\ref{eqn:face.integrated.dp} exactly.

In the optically-thick limit of particular interest, Eq.~\ref{eqn:face.integrated.dp} is approximately $\approx c^{-1}\, \int_{\Va,\,\Omega_{ab}} d^{3}{\bf x}\,d\Omega\,{\rho\,\kappa_{\nu} I_{\nu}\,\hat{\Omega}}$.  In discrete (Monte-Carlo) methods, this becomes 
$\Delta {\bf p}_{ab} = c^{-1}\,\sum_{j}\,\Theta_{ab}({\bf x}_{\rm abs},\,\hat{\Omega}_{j})\,E_{\gamma}^{j,\,{\rm abs}}\,\hat{\Omega}_{j}$.

Since the momentum flux is defined now at {\em faces}, rather than cell centers, it can be incorporated either as a simple operator-split flux (i.e.\ cell ``$b$'' receives a momentum $\Delta {\bf p}_{ab}$ from cell ``$a$''), or as a force in the Riemann problem or momentum equation solved between cells. Note that in general $| \pdotab | \ne |\pdotba |$, so the {\em net} momentum flux between cells, $\pdotab + \pdotba$, is what will ultimately matter. But for a single source in our example problem, $\pdotba = {\bf 0}$, so the force is purely ``outwards'' from the source (as it should be), while for a perfectly symmetric or homogeneous source distribution, $\pdotba=-\pdotab$ and the net forces correctly vanish. 

Return to our test problem: it is easy to verify that this produces an outward force from the central cell. The momentum flux into each neighbor $b$ follows the exact solution (assuming a spherical shell propagates out from the center), averaged over the face. While initially the solution cannot, of course, be exactly spherical (owing to the grid geometry), it will rapidly converge to such as the shell expands.

%\vspace{-0.5cm}
%\subsubsection{Implementation in Ray-Tracing or Monte Carlo Methods}

%, this integral can be obtained by simply taking each absorption ``event'' and adding the momentum to the nearest face along the continuation of the ray/trajectory, i.e.: 
%\begin{align}
%\Delta {\bf p}_{ab} = c^{-1}\,\sum_{j}\,\Theta_{ab}({\bf x},\,\hat{\Omega}_{j})\,E_{\gamma}^{j,\,{\rm abs}}\,\hat{\Omega}_{j}
%\end{align}
%where $\Theta_{ab}({\bf x},\,\hat{\Omega}_{j}) = 1$ if the face ${\bf A}_{ab}$ is the ``nearest intercepted face'' (i.e.\ if ${\bf A}_{ab}$ is the first face encountered, along the ray originating at ${\bf x}$ in direction $\hat{\Omega}_{j}$), and $=0$ otherwise.

%Although approximate ray-tracing or Monte-Carlo methods usually sample only a subset of all possible angles $\hat{\Omega}$, the error from evaluating this discretely only along the sampled $\hat{\Omega}$ will usually be small, so long as a sufficient number of rays or photon packets are used to sample the different cardinal directions (usually a pre-requisite in such methods).

\vspace{-0.5cm}
\subsubsection{Implementation in Moments-Based Methods: Additional Complications and Errors}
\label{sec:moments.additional.problems}

In moments-based methods, two additional, independent and very important errors arise. 

\begin{enumerate}

\item Because the ``mesh'' on which radiative properties are computed is identical to the fluid mesh, the flux ${\bf F}_{\nu}({\bf x}) = \langle {\bf F}_{\nu} \rangle_{a}$ is single valued within a given cell. So we must insert an {\em explicit} model for ${\bf F}_{\nu}({\bf x})$ around each source. Since the error here, where important ($\Delta x \gg \mfp$), is dominated by the cell $a$ surrounding the source, this can be accomplished during the ``photon deposition'' step. Recall, in moments methods photons must be ``deposited'' onto the grid around a source, normally via a scalar kernel/weight function: e.g.\ $(\Delta E_{\nu})_{ab} = L_{a}\,\Delta t_{a}\,\,\omega_{ab}$ where $\sum_{b}\omega_{ab}=1$. So {\em during this step}, one can {\em explicitly} calculate the integral in Eq.~\ref{eqn:face.integrated.dp} with ${\bf F}_{\nu}$ from each source, integrated over domain $\Va,\,\Omega_{ab}$ within cell $a$ towards face $ab$. A detailed example of how to do this numerically is given in Appendix~\ref{sec:appendix.face.construction}. Outside the cell hosting a source, it is a much smaller error to simply adopt the cell-centered $\langle {\bf F}_{\nu} \rangle$ (if $\Delta x \ll \mfp$, the solution is converged, if $\Delta x \gg \mfp$, the flux outside the origin cell $a$ is negligible).

\item In moments methods, the flux ${\bf F}$ does not directly follow from $I_{\nu}$, but is sourced by the {\em numerical gradient} of $e_{\nu}$. This means that even outside the ``origin'' cell, it requires several resolution elements to resolve any gradient in $e_{\nu}$ (consider, at the origin, the numerical gradient must vanish if the source is isotropic). So even when $\Delta x \lesssim \mfp$, essentially no photon momentum can be transferred to the gas within the central $\sim 10$ cells in any direction because the {\em numerical} gradient is smoothed by a finite discrete ``kernel.'' This is true even for higher-order gradient estimators in regular Cartesian-mesh codes (see Fig.~B1 in \citealt{rosdahl:2015.galaxies.shine.rad.hydro}). This means a {\em mass} $\sim (4\pi/3)\,(10\,\Delta x)^{3}\,\rho \sim 4200\,\Delta m$ has its momentum under-estimated, which makes the resolution criterion in Eqs.~\ref{eqn:spatial.res}-\ref{eqn:mass.res} much more challenging (the mass resolution must be at least $>4000$ times better, or spatial resolution $>20$ times better, to converge). One fix to this is to extend the explicit integration method for ${\bf F}_{\nu}$ in (i) above, to a $\sim 10$-cell radius around each source -- essentially, performing a ``mini ray-trace'' in each $r \sim 10\,\Delta x$ sphere around each source. This is the most accurate method, but is computationally much more complex. A much simpler, albeit significantly less accurate, fix is proposed and adopted by \citet{rosdahl:2015.galaxies.shine.rad.hydro}: simply replace ${\bf F}_{\nu} \rightarrow e_{\nu}\,c\,\hat{\bf F}_{\nu}$ whenever calculating the photon momentum/radiation pressure terms. This is exact in the single-source, optically-thin limit; however if there are multiple sources with intersecting rays this can over-estimate the true momentum transfer (although the error is usually small unless the sources should exactly cancel). 

\end{enumerate}

%\item As noted in \S~\ref{sec:moments.overview}, by {\em assumption} in such methods $I_{\nu}(\hat{\Omega},\,{\bf x},\,t)$ is single-valued at each point ${\bf x}$ (non-zero only in the direction $\hat{\bf F}({\bf x})$). So Eq.~\ref{eqn:face.integrated.dp} collapses to Eq.~\ref{eqn:face.integrated.flux} ($\pdotab \rightarrow c^{-1}\,\int_{{\rm Vol}_{a},\,\Omega_{ab}} {\rho\,\kappa_{\nu}\,{\bf F}_{\nu}}\,d^{3}{\bf x}$). This means one can never converge to the correct solution for certain physical situations (e.g.\ two nearby, optically-thin sources producing overlapping rays with opposite direction), regardless of resolution. However this is a fundamental limitation of the moment methods, not unique to the coupling of momentum back to the grid (our focus here).

%\item More problematic for our purposes, since in  moments methods the ``mesh'' on which radiative properties are computed is identical to the fluid mesh, the flux ${\bf F}_{\nu}({\bf x}) = \langle {\bf F}_{\nu} \rangle_{a}$ is single valued within a given cell.

\vspace{-0.5cm}
\subsection{Numerical Example}
\label{sec:numerical.test}

Fig.~\ref{fig:acc.calc} demonstrates the errors and their fixes here with a variant of the simple test problem in Fig.~\ref{fig:single.scattering.demo}. We place a single, isotropic, monochromatic point source of constant luminosity $L$ randomly on an effectively infinite grid of uniform cartesian cells of size $\Delta x$ with constant density $\rho$ and opacity $\kappa$, and calculate the equilibrium flux ${\bf F}$ at all points (assuming the light-crossing time is much shorter than all other timescales in the problem) according to the labeled RHD method. Again we stress the gas properties are ``frozen'' - we are just solving the RTE. We then calculate the momentum flux $\dot{\bf p}$ {\em either} according to the cell-integrated or face-integrated approach. For the moments methods our ``face-integrated'' method also implements the additional fixes from \S~\ref{sec:moments.additional.problems}. We calculate the total ``outward''/radial momentum flux from the source, $\dot{p}_{\hat{r}}^{\rm tot} = \sum_{ab} (\dot{\bf p})_{a} \cdot \hat{r}_{a}$ where $\hat{r}_{a}$ is the unit vector pointing from source to the cell center in the cell-integrated case; for the face-integrated case we sum over $(\dot{\bf p})_{ab}\cdot \hat{r}_{ab}$ where $\hat{r}_{ab}$ points to the face-center. This has the exact solution $\dot{p}_{\hat{r}}^{\rm tot} = L/c$. We compare $\dot{p}_{\hat{r}}^{\rm tot}$ for each method, as a function of $\Delta x/\mfp$ -- in these units, the absolute units of the problem scale out completely. 

As expected, when the momentum is deposited in a ``cell-integrated'' fashion, even {\em exact} ray-tracing or MC methods severely suppress the momentum unless $\Delta x \ll 0.5\,\mfp$. The ``cell-integrated,'' ``default'' moments (M1/FLD) methods fare even worse, requiring $\Delta x \ll 0.05\,\mfp$. With ``face-integrated'' methods, however, the errors are radically reduced, and depend only very weakly on resolution (owing to geometric effects discussed above). 

At low resolution ($\Delta x/\mfp \rightarrow \infty$), even with an exact RT solution, the face-integrated methods converge to a coupled radial momentum $\dot{p}_{\hat{r}}^{\rm tot} \rightarrow 0.84\, L/c$, not $L/c$ (i.e.\ the total {\em radial} momentum is $\sim 16\%$ lower than the converged solution). This is a real geometric effect.\footnote{In fact the rectilinear Cartesian mesh we adopt gives nearly the worst-case scenario for this geometric effect. If we instead assume a glass configuration of mesh-generating points with a Voronoi tesselation or the mesh-free MFM/MFV methods from \citet{hopkins:gizmo} used to calculate the faces, the larger number of faces (cells are closer to regular polyhedra with $\sim 16$ faces) means the mean face-integrated momentum as $\Delta x/\mfp \rightarrow \infty$ is $\dot{p}_{\hat{r}}^{\rm tot} \rightarrow 0.94\, L/c$.} Imagine a perfectly-resolved spherical shell emerges from $a$ with total radial momentum $=L/c$ -- but upon entering cell $b$, we must {\em integrate} the total momentum entering $b$, which means averaging the momentum of the parts of the shell crossing ${\bf A}_{ab}$ (producing some cancellation of the momentum components transverse to the face). So the solution is still ``exact'' in that it reflects the converged solution, ``averaged down'' to the grid scale after propagating through each cell. More important, this difference is small and constant (while cell-integrated methods have $\dot{p}_{\hat{r}}^{\rm tot}$ decreasing exponentially as $\Delta x/\mfp \rightarrow \infty$).

For this simple test problem, approximate-ray (LEBRON/ART) methods give the same result as exact ray-tracing/MC methods. Moments methods always produce larger errors for the reasons in \S~\ref{sec:moments.additional.problems} (because we allow the radiation field to reach equilibrium, FLD and M1 are identical here). Even with the face-integrated method and fixes in \S~\ref{sec:moments.additional.problems}, moments methods are not perfect: at $\Delta x \ll \mfp$ they converge (much faster than with cell-integrated coupling), and at $\Delta x \gg \mfp$ the solution occurs entirely ``within one cell'' so they are identical to the exact methods, but when $\Delta x \sim \mfp$, the gradient errors noted in \S~\ref{sec:moments.additional.problems} somewhat suppress the coupled momentum.

\vspace{-0.5cm}
\section{Extension to Multiple-Scattering}
\label{sec:multiple.scattering}

Briefly, consider multiple-scattering: photons are absorbed and re-emitted (or scattered) many times before escaping/being destroyed. Where this is of interest, opacities are usually low $\kappa \sim \kappa_{0}\,{\rm cm^{2}\,g^{-1}}$: e.g.\ in IR dust re-processing ($\kappa_{0}\sim 1-10$), free-electron scattering ($\kappa_{0}\sim 0.4$), X-ray metal-line absorption ($\kappa_{0}\sim 2$ at the Fe edge). So resolution requirements are less extreme: $\Delta x \ll \mfp \sim 20\,\kappa_{0}^{-1}\,n_{4}^{-1}\,{\rm pc}$ or $\Delta m \ll \mfp^{2}/\kappa_{\nu} \sim 2\times10^{6}\,\kappa_{0}^{-3}\,n_{4}^{-2}\,\msun$. But even these are not always met; moreover multiple-scattering can also be important in resonance lines where $\kappa$ is much larger. 

It is trivial to see that in this limit once again, a ``cell-integrated'' coupling of photon momentum to gas (Eq.~\ref{eqn:cell.integrated.dp}) gives essentially vanishing radiation pressure $(\dot{\bf p}_{\nu})_{ab} \rightarrow {\bf 0}$ in the cell around a source when $\Delta x \gg \mfp$ (precisely the opposite of the correct behavior). But it is also easy to verify that the ``face-integrated'' coupling (Eq.~\ref{eqn:face.integrated.dp}) gives the correct solution in this limit. 

We can repeat the numerical experiment in Fig.~\ref{fig:acc.calc}, for example, but instead of single-scattering assume multiple-scattering with a gray (constant) opacity $\kappa$, take $\mfp\rightarrow 0$ compared to resolved scales, and assume perfect re-emission (all absorbed energy is re-radiated). In that case the exact solution is now $\dot{p}_{\hat{r}}^{\rm tot}(<r) = \tau(<r)\,L/c$, where $\dot{p}_{\hat{r}}^{\rm tot}(<r)$ is the total radial momentum flux integrated over all cells within a radius $r$, and $\tau(<r) = \rho\,\kappa\,r$ is the optical depth out to that radius. If we then compare $\dot{p}_{\hat{r}}^{\rm tot}(<r)$ to the correct solution $\tau(<r)\,L/c$, as a function now of $\Delta x / r$ (since $\mfp\rightarrow0$ by assumption), we obtain essentially identical results to those shown in Fig.~\ref{fig:acc.calc} for single-scattering.

\begin{figure}
\centering
    \plotonesize{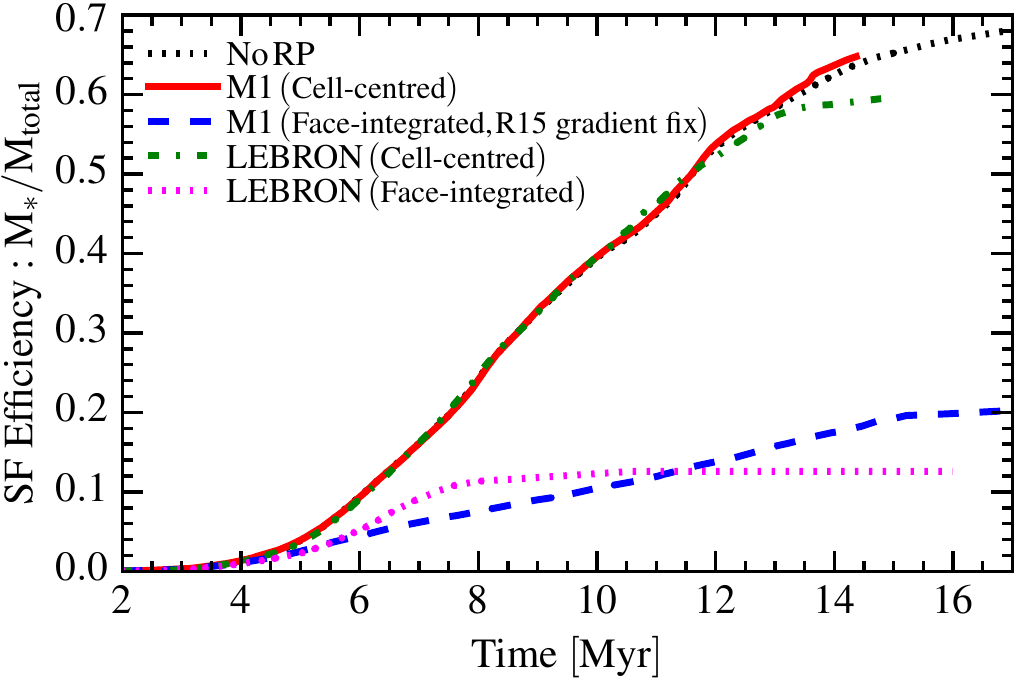}{0.95}
    %\plotsidesize{rtmom_acc_faceint.pdf}{0.49}
    \vspace{-0.2cm}
    \caption{Example problem (see \S~\ref{sec:test.problem}), illustrating effects of the errors discussed here. We simulate a single GMC collapsing and forming a star cluster, including self-gravity, star formation, and single-scattering RP (each star/sink, once formed, is assigned a constant mono-chromatic luminosity used to solve the RTE assuming single-scattering). We compare two approximate methods to solve the RTE, chosen to have opposite regimes of validity: (1) the ray-based LEBRON (exact in the optically-thin single-scattering limit), and (2) the moments-based M1 (exact in the optically-thick {\em multiple}-scattering limit). For each we compare cell-centred and face-integrated formulations. We plot the stellar mass (as a fraction of total cloud mass) versus time. Without RP, most of the mass turns into stars (no other feedback, e.g.\ photo-ionization heating or SNe, is included here). With both RT methods, the cell-centred formulations suppress the RP strongly (it has almost no effect), while the more accurate face-integrated couplings predict a large effect of RP suppressing SF.\vspace{-0.5cm}
    \label{fig:example.starcluster}}
\end{figure}

\vspace{-0.5cm}
\section{Effects in an Example Problem: Star Cluster Formation}
\label{sec:test.problem}

To explore this in an astrophysical context, we consider as an example a simulation of star cluster formation in an individual GMC following \citet{grudic:sfe.cluster.form.surface.density}. The simulations use the code GIZMO \citep{hopkins:gizmo},\footnote{\gizmourl} with the meshless-finite-mass (MFM) Godunov MHD solver, and fully-adaptive/Lagrangian force-softening. We initialize a solar-metallicity cloud of mass $10^{6}\,\msun$ and size $50\,{\rm pc}$, with intentionally low resolution (fixed $\Delta m=8\,\msun$), and an initially turbulent velocity and magnetic field spectrum with virial parameter of unity and mean plasma $\beta=100$, and evolve it including self-gravity, ideal MHD, radiative cooling (with the cooling curves from \citealt{hopkins:fire2.methods} from $10-10^{10}$\,K, assuming a universal Milky-Way like radiation field for heating/cooling), and star formation into sink particles in locally self-gravitating, Jeans-unstable, converging flows \citep[all details in][]{grudic:sfe.cluster.form.surface.density}. While the simulations in \citet{grudic:sfe.cluster.form.surface.density} included stellar mass-loss, SNe (Ia \&\ II), and multi-band RHD with single and multi-scattering RP \&\ photo-heating, here we disable these terms and consider {\em only} monochromatic single-scattering RP (to clearly isolate the effects of interest). To further simplify, each star particle (once formed) is assigned a constant light-to-mass ratio $L/M = 1100\,L_{\sun}/M_{\sun}$ and we adopt a constant $\kappa=2000\,{\rm cm^{2}\,g^{-1}}$ (approximately appropriate for near-UV luminosities \&\ dust opacities around young populations). 

We solve RHD using one of two approximate methods: (1) the ray-based LEBRON (for implementation details, see \citealt{hopkins:fire2.methods}), or (2) moments-based M1 (following \citealt{rosdahl:2013.m1.ramses}). For each, we consider both ``cell-integrated'' and ``face-integrated'' implementations as Figs.~\ref{fig:single.scattering.demo}-\ref{fig:acc.calc} (with the ``face-integrated'' M1 also including the \citealt{rosdahl:2015.galaxies.shine.rad.hydro} fix to the gradient error discussed in \S~\ref{sec:moments.additional.problems}). We also run a test with no radiation for comparison. We emphasize that both of these methods will have large errors here: LEBRON is exact in the optically-thin limit, M1 in the optically-thick multiple-scattering limit, so neither solves the RTE exactly in this problem where there are both thin \&\ thick single-scattering regimes. So since their errors and regimes of applicability are quite different, it is useful to consider how both methods are influenced by these errors.

Fig.~\ref{fig:example.starcluster} plots the ``star formation efficiency'' (SFE; fraction of the mass turned into stars): over a few dynamical times, gas collapses and rapidly turns into stars until it is exhausted or expelled by RP. In both cell-centered runs, the RP is strongly suppressed at this resolution and has no effect on the SFE. In both face-centered runs, the SFE is suppressed by a significant factor -- i.e.\ RP has a dramatic effect on the cluster evolution.\footnote{The Fig.~\ref{fig:example.starcluster} runs with ``fixed'' schemes give roughly the expected result from simple analytic arguments: comparing the strength of the total RP force $\sim L/c \sim (L/M)\,M_{\ast}/c$ to gravity $\sim G\,M_{\rm cloud}^{2}/R^{2}$ shows that RP should over-whelm gravity on the cloud scale when $M_{\ast} \gtrsim 0.1\,M_{\rm cloud}$.} This may explain several apparently discrepant results in the literature. For example, the default \citet{grudic:sfe.cluster.form.surface.density} simulations adopted the ``face-integrated'' LEBRON method, while a similar study by \citet{raskutti:2016.m1.cloud.sims} used the ``cell-integrated'' (and otherwise ``un-corrected'') M1 method. As expected from our test here, \citet{raskutti:2016.m1.cloud.sims} find an order-of-magnitude higher SFE (much weaker effects of radiation pressure), for otherwise similar clouds.

It is worth noting that the errors here can be ``hidden'' in numerical studies which include e.g.\ {\em both} RP and other feedback (e.g.\ photo-ionization or stellar winds or SNe). Consider: if some other (non-RP) physics first creates a low-opacity ``bubble'' around the source (e.g.\ stellar winds sweeping gas into a shell surrounding a very low-density cavity, or SNe destroying dust, or photo-ionization generating a Stromgren sphere which eliminates the neutral hydrogen opacity inside the sphere and/or pushes gas again into a shell), then one can have a mean free path {\em immediately around the source} which becomes resolved ($\Delta x \ll \mfp$), even though when the photons eventually encounter a distant high-opacity ``shell'' (or neutral gas or dust) they are absorbed in a thin layer ($\Delta x \gg \mfp$ ``in the shell''). In that regime, as long as the shell radius is well-resolved, the errors here are not large (the momenta do not cancel inside a single cell) -- so simulations like the GMC study in Fig.~\ref{fig:example.starcluster} which attempt to be more ``full physics'' may have the inadvertent benefit of enabling better resolution and therefore coupling of the RP forces (making the differences less stark). However, even in these cases, the {\em early} RP effects (before the shell/Stromgren sphere reaches large radii $r \gg \Delta x$) will still be severely suppressed (if $\Delta x \gg \mfp$). Moreover, if one takes one of these simulations and ``turns off'' the other sources of feedback, then RP alone would be dramatically suppressed, leading to the incorrect conclusion that it is un-important. And more generally, one cannot always rely on these secondary mechanisms to ensure the RP forces ``become resolved.'' In sufficiently dense clouds, photo-ionization will not generate an expanding bubble or large/well-resolved HII regions, and in galaxy-scale or cosmological-scale simulations even quite large individual HII regions are usually un-resolved. And in different physical circumstances, e.g.\ outflows from AGN, it is not obvious whether any other process necessarily ``pre-generates'' a cavity for the RP to act within.

\vspace{-0.5cm}
\section{Conclusions}
\label{sec:discussion}

We have shown that ``cell-integrated'' coupling of absorbed photon momentum (radiation pressure) in radiation-hydrodynamics treatments -- the most common approach used in the literature -- severely under-estimates the true momentum flux around sources, unless the photon mean-free-paths are well-resolved in the {\em hydrodynamic} grid (fluid spatial resolution $\Delta x \ll \mfp$ or mass resolution $\Delta m \ll \mfp^{2}/\kappa$). But the required resolution is often impossible -- for example, in simulations of star or galaxy formation, proper treatment of the UV \&\ ionizing photons (most of the single-scattering radiation pressure) formally requires mass resolution $\Delta m \ll  10^{-13}\,\msun$ (or $\ll  10^{-17}\,\msun$ in FLD/M1 ``moments'' methods)! 

Fortunately, we show that adopting a ``face-integrated'' coupling -- in which the momentum flux is integrated towards each hydrodynamic face, and treated as part of the usual fluxes, instead of being integrated over the entire cell volume and added to the cell-centered momentum -- resolves these errors. We show that this produces good convergence (provided the radiative transfer equation is properly solved) even when the gas grid fails to resolve the photon mean free paths ($\Delta x \gg \mfp$). 

We stress that the errors identified here, and their fixes, are {\em independent} of the radiation methods, even if the RTE is solved {\em perfectly} (with e.g.\ exact Monte-Carlo or ray-tracing methods). The error arises not from the solution of the RTE, but from {\em how the absorbed photon momentum is assigned to the gas}. As such the errors and their fixes are extremely general. We do also identify some additional (related but distinct) errors in common implementations unique to moments-based methods (FLD/OTVET/M1) for RHD and discuss potential fixes. We also demonstrate that the errors and fixes are qualitatively independent of the hydrodynamic method -- we outline how to implement these for both grid-based codes and mesh-free (e.g.\ SPH) methods. 

We show that the erroneous formulation can fail entirely to capture the effects of radiation pressure (under-estimating the true photon momentum by orders-of-magnitude) when $\Delta x \ll \mfp$. We illustrate the effects of this in a fully non-linear example of state-of-the-art simulations of star cluster formation, where we show an improper cell-integrated treatment of the single-scattering photon-momentum leads to incorrect conclusions. Specifically (in the test chosen), with a correct treatment of the radiation pressure (in either M1 or ray-based methods), radiation pressure from massive stars rapidly disrupts the star-forming GMC and greatly suppresses the star formation efficiency. With an incorrect (cell-integrated) treatment (again in both M1 and ray-based methods) the radiation pressure (erroneously) does very little to the cloud. Obviously similar extensions apply to any simulations of unresolved point-like sources, for example most simulations of the effects of AGN radiation on regions larger than the emitting disk (e.g.\ the broad-line or torus or narrow-line regions). 

We stress that simple ``resolution tests'' (running the same ICs with increasing resolution) applied to this problem will not reveal the problem with cell-integrated approaches, because the behavior will change sharply (and converge to the correct solution) only when the mass resolution $\Delta m$ becomes smaller than the ludicrously-small value above (which would require $\gtrsim 10^{16}$ particles in our example star-cluster simulation). At achievable resolution, one will instead see ``false convergence'' until this threshold is reached.

\vspace{-0.7cm}
\acknowledgments 
Support for PFH and MYG was provided by an Alfred P. Sloan Research Fellowship, NSF Collaborative Research Grant \#1715847 and CAREER grant \#1455342. Numerical calculations were run on the Caltech compute cluster ``Wheeler,'' allocations from XSEDE TG-AST130039 and PRAC NSF.1713353 supported by the NSF, and NASA HEC SMD-16-7592. \\

\vspace{-0.2cm}
\bibliography{/Users/phopkins/Dropbox/Public/ms}

\begin{thebibliography}{31}
\expandafter\ifx\csname natexlab\endcsname\relax\def\natexlab#1{#1}\fi

\bibitem[{{Abel} \& {Wandelt}(2002)}]{abel:2002.adaptive.ray.tracing}
{Abel}, T., \& {Wandelt}, B.~D. 2002, \mnras, 330, L53

\bibitem[{{Bate}(2012)}]{bate:2012.rmhd.sims}
{Bate}, M.~R. 2012, \mnras, 419, 3115

\bibitem[{{Buntemeyer} {et~al.}(2016){Buntemeyer}, {Banerjee}, {Peters},
  {Klassen}, \& {Pudritz}}]{buntemeyer:2016.art}
{Buntemeyer}, L., {Banerjee}, R., {Peters}, T., {Klassen}, M., \& {Pudritz},
  R.~E. 2016, New Astronomy, 43, 49

\bibitem[{{Davis} {et~al.}(2014){Davis}, {Jiang}, {Stone}, \&
  {Murray}}]{davis:2014.rad.pressure.outflows}
{Davis}, S.~W., {Jiang}, Y.-F., {Stone}, J.~M., \& {Murray}, N. 2014, \apj,
  796, 107

\bibitem[{{Davis} {et~al.}(2012){Davis}, {Stone}, \&
  {Jiang}}]{davis:2012.rhd.short.characteristics}
{Davis}, S.~W., {Stone}, J.~M., \& {Jiang}, Y.-F. 2012, \apjs, 199, 9

\bibitem[{{Fleck} \& {Cummings}(1971)}]{fleck:1971.implicit.monte.carlo}
{Fleck}, Jr., J.~A., \& {Cummings}, J.~D. 1971, Journal of Computational
  Physics, 8, 313

\bibitem[{{Foucart}(2018)}]{foucart:2018.pic.closure}
{Foucart}, F. 2018, \mnras

\bibitem[{{Gnedin} \& {Abel}(2001)}]{gnedin.abel.2001:otvet}
{Gnedin}, N.~Y., \& {Abel}, T. 2001, New Astronomy, 6, 437

\bibitem[{{Gonz{\'a}lez} {et~al.}(2015){Gonz{\'a}lez}, {Vaytet}, {Commer{\c
  c}on}, \& {Masson}}]{gonzalez:2015.fld.rhd}
{Gonz{\'a}lez}, M., {Vaytet}, N., {Commer{\c c}on}, B., \& {Masson}, J. 2015,
  \aap, 578, A12

\bibitem[{{Grudi{\'c}} {et~al.}(2018){Grudi{\'c}}, {Hopkins},
  {Faucher-Gigu{\`e}re}, {Quataert}, {Murray}, \& {Kere{\v
  s}}}]{grudic:sfe.cluster.form.surface.density}
{Grudi{\'c}}, M.~Y., {Hopkins}, P.~F., {Faucher-Gigu{\`e}re}, C.-A.,
  {Quataert}, E., {Murray}, N., \& {Kere{\v s}}, D. 2018, \mnras, 475, 3511

\bibitem[{{Hopkins}(2015)}]{hopkins:gizmo}
{Hopkins}, P.~F. 2015, \mnras, 450, 53

\bibitem[{{Hopkins} {et~al.}(2014){Hopkins}, {Keres}, {Onorbe},
  {Faucher-Giguere}, {Quataert}, {Murray}, \& {Bullock}}]{hopkins:2013.fire}
{Hopkins}, P.~F., {Keres}, D., {Onorbe}, J., {Faucher-Giguere}, C.-A.,
  {Quataert}, E., {Murray}, N., \& {Bullock}, J.~S. 2014, \mnras, 445, 581

\bibitem[{{Hopkins} {et~al.}(2011){Hopkins}, {Quataert}, \&
  {Murray}}]{hopkins:rad.pressure.sf.fb}
{Hopkins}, P.~F., {Quataert}, E., \& {Murray}, N. 2011, \mnras, 417, 950

\bibitem[{{Hopkins} {et~al.}(2018{\natexlab{a}}){Hopkins}, {Wetzel}, {Kere{\v
  s}}, {Faucher-Gigu{\`e}re}, {Quataert}, {Boylan-Kolchin}, {Murray},
  {Hayward}, {Garrison-Kimmel}, {Hummels}, {Feldmann}, {Torrey}, {Ma},
  {Angl{\'e}s-Alc{\'a}zar}, {Su}, {Orr}, {Schmitz}, {Escala}, {Sanderson},
  {Grudi{\'c}}, {Hafen}, {Kim}, {Fitts}, {Bullock}, {Wheeler}, {Chan},
  {Elbert}, \& {Narayanan}}]{hopkins:fire2.methods}
{Hopkins}, P.~F., {et~al.} 2018{\natexlab{a}}, \mnras, 480, 800

\bibitem[{{Hopkins} {et~al.}(2018{\natexlab{b}}){Hopkins}, {Wetzel}, {Kere{\v
  s}}, {Faucher-Gigu{\`e}re}, {Quataert}, {Boylan-Kolchin}, {Murray},
  {Hayward}, \& {El-Badry}}]{hopkins:sne.methods}
---. 2018{\natexlab{b}}, \mnras, 477, 1578

\bibitem[{{Kim} {et~al.}(2017){Kim}, {Kim}, {Ostriker}, \&
  {Skinner}}]{kim:2017.art.uv.starclusters}
{Kim}, J.-G., {Kim}, W.-T., {Ostriker}, E.~C., \& {Skinner}, M.~A. 2017, \apj,
  851, 93

\bibitem[{{Kolb} {et~al.}(2013){Kolb}, {Stute}, {Kley}, \&
  {Mignone}}]{pluto:2013.fld.implicit}
{Kolb}, S.~M., {Stute}, M., {Kley}, W., \& {Mignone}, A. 2013, \aap, 559, A80

\bibitem[{{Kuiper} {et~al.}(2012){Kuiper}, {Klahr}, {Beuther}, \&
  {Henning}}]{kuiper:2012.rad.pressure.outflow.vs.rt.method}
{Kuiper}, R., {Klahr}, H., {Beuther}, H., \& {Henning}, T. 2012, \aap, 537,
  A122

\bibitem[{{Levermore}(1984)}]{levermore:1984.FLD.M1}
{Levermore}, C.~D. 1984, Journal of Quantitative Spectroscopy and Radiative
  Transfer, 31, 149

\bibitem[{{Levermore} \& {Pomraning}(1981)}]{levermore:1981.fld}
{Levermore}, C.~D., \& {Pomraning}, G.~C. 1981, \apj, 248, 321

\bibitem[{{Lowrie} {et~al.}(1999){Lowrie}, {Morel}, \&
  {Hittinger}}]{lowrie:1999.radiation.hydro.coupling}
{Lowrie}, R.~B., {Morel}, J.~E., \& {Hittinger}, J.~A. 1999, \apj, 521, 432

\bibitem[{{Mihalas} \& {Mihalas}(1984)}]{mihalas:1984oup..book.....M}
{Mihalas}, D., \& {Mihalas}, B.~W. 1984, {Foundations of radiation
  hydrodynamics} (New York, Oxford University Press, 731 p.)

\bibitem[{{Olson} \& {Kunasz}(1987)}]{olson:1987.short.char.rhd}
{Olson}, G.~L., \& {Kunasz}, P.~B. 1987, Journal of Quantitative Spectroscopy
  and Radiative Transfer, 38, 325

\bibitem[{{Raskutti} {et~al.}(2016){Raskutti}, {Ostriker}, \&
  {Skinner}}]{raskutti:2016.m1.cloud.sims}
{Raskutti}, S., {Ostriker}, E.~C., \& {Skinner}, M.~A. 2016, \mnras, submitted,
  arXiv:1608.04469

\bibitem[{{Rosdahl} {et~al.}(2013){Rosdahl}, {Blaizot}, {Aubert}, {Stranex}, \&
  {Teyssier}}]{rosdahl:2013.m1.ramses}
{Rosdahl}, J., {Blaizot}, J., {Aubert}, D., {Stranex}, T., \& {Teyssier}, R.
  2013, \mnras, 436, 2188

\bibitem[{{Rosdahl} {et~al.}(2015){Rosdahl}, {Schaye}, {Teyssier}, \&
  {Agertz}}]{rosdahl:2015.galaxies.shine.rad.hydro}
{Rosdahl}, J., {Schaye}, J., {Teyssier}, R., \& {Agertz}, O. 2015, \mnras, 451,
  34

\bibitem[{{Rosen} {et~al.}(2017){Rosen}, {Krumholz}, {Oishi}, {Lee}, \&
  {Klein}}]{rosen:2017.harm.rhd}
{Rosen}, A.~L., {Krumholz}, M.~R., {Oishi}, J.~S., {Lee}, A.~T., \& {Klein},
  R.~I. 2017, Journal of Computational Physics, 330, 924

\bibitem[{{Roth} \& {Kasen}(2015)}]{roth:2015.implicit.mc}
{Roth}, N., \& {Kasen}, D. 2015, \apjs, 217, 9

\bibitem[{{Tominaga} {et~al.}(2015){Tominaga}, {Shibata}, \&
  {Blinnikov}}]{tominaga:2015.spherical.harmonic.rhd}
{Tominaga}, N., {Shibata}, S., \& {Blinnikov}, S.~I. 2015, \apjs, 219, 38

\bibitem[{{Wise} {et~al.}(2012){Wise}, {Abel}, {Turk}, {Norman}, \&
  {Smith}}]{wise:2012.rad.pressure.effects}
{Wise}, J.~H., {Abel}, T., {Turk}, M.~J., {Norman}, M.~L., \& {Smith}, B.~D.
  2012, \mnras, 427, 311

\bibitem[{{Zhang} \& {Davis}(2017)}]{zhang:2017.rhd.dusty.winds}
{Zhang}, D., \& {Davis}, S.~W. 2017, \apj, 839, 54

\end{thebibliography}
%\bibliography{ms_trimmed}

\begin{appendix}

\vspace{-0.5cm}
\section{Effective Face Construction and Coupling: Example Methods}
\label{sec:appendix.face.construction}

\begin{figure*}
    \plotsidesize{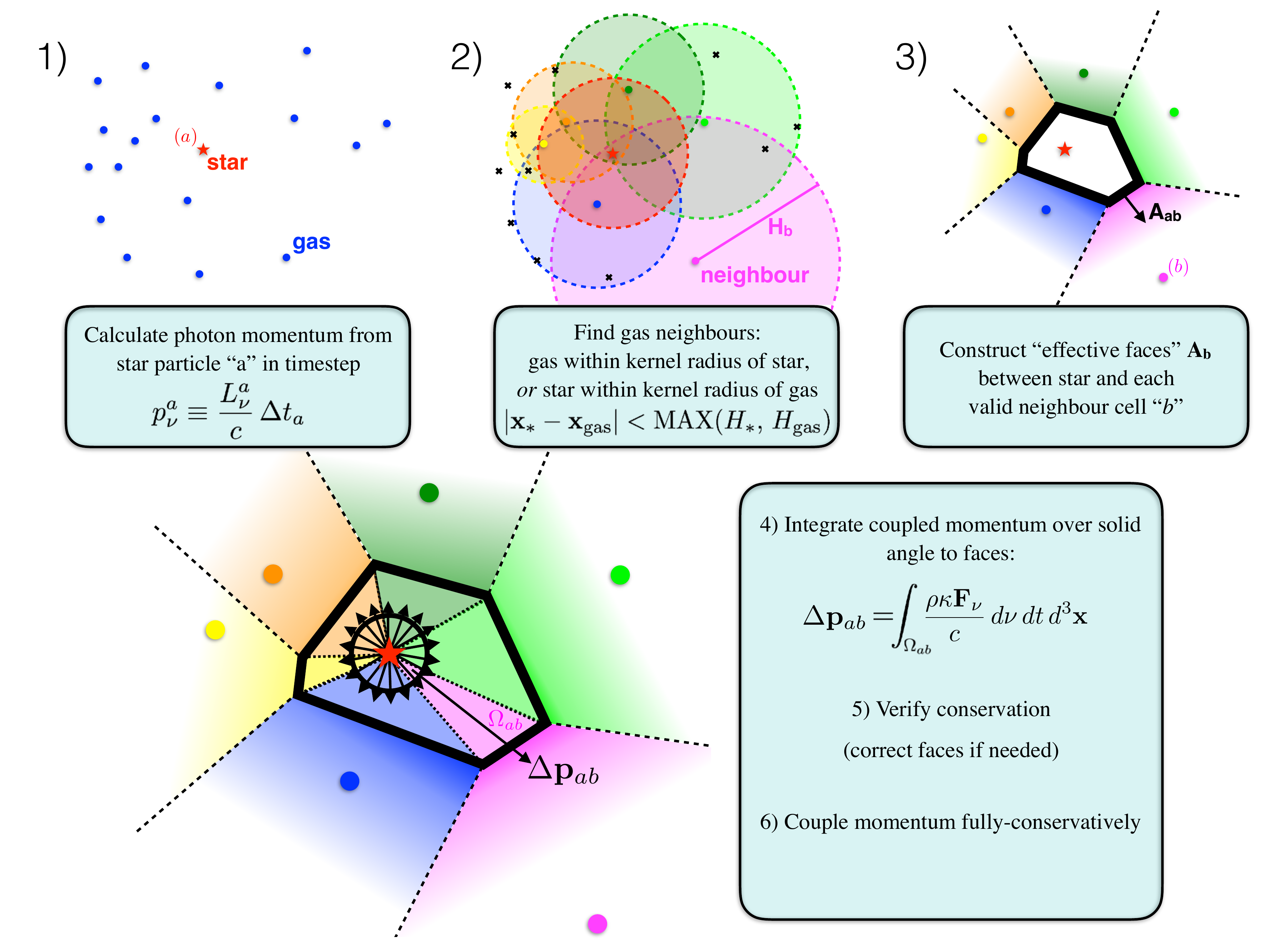}{0.99}
    \vspace{-0.25cm}
    \caption{Cartoon illustrating a ``face-integrated'' numerical algorithm for coupling radiative feedback in arbitrary mesh or particle-based codes (\S~\ref{sec:appendix.face.construction}). 
    {\bf (1)} Determine whether a source $a$ (e.g.\ a star or black hole) is active in a given timestep, and calculate its luminosity in each waveband. 
    {\bf (2)} Identify valid interacting neighbors $b$ for the source (fluid elements within a search radius $H_{a}$ from the source, or for which the source is within their search radius $H_{b}$), taking care to cover all directions from the source. 
    {\bf (3)} Construct the ``effective faces'' ${\bf A}_{ab}$ of the interacting fluid elements, as seen by the source, via e.g.\ a Voronoi tessellation or kernel-volume decomposition.
    {\bf (4)} Integrate the absorbed photon momentum over the solid angle subtended by each face ${\bf A}_{ab}$ (i.e.\ in the domain $\Omega_{ab}$), according to Eq.~\ref{eqn:face.integrated.dp}, to determine the momentum flux $(\dot{\bf p}_{\nu})_{ab}$ associated with each face.
    {\bf (5)} Verify the fluxes maintain machine-accurate momentum conservation and statistical isotropy; if not, correct the faces to ensure this is true (implicit in the algorithm here). 
    {\bf (6)} Couple the momentum flux at faces in a fully-conservative manner, either as an operator-split force or added to the force on the face from the Riemann problem: e.g.\ ${\bf p}_{b} = {\bf p}_{b}(t) + \sum_{\rm sources}\sum_{\rm faces} \Delta {\bf p}_{ab}$ with $\Delta {\bf p}_{ab} = \int_{\Delta \nu}\int _{\Delta t} (\dot{\bf p}_{\nu})_{ab}\,d\nu\,dt$. 
    \label{fig:faces.cartoon}}
\end{figure*}

As discussed in the text, our face-integrated methods require defining effective faces ${\bf A}_{ab}$ around any source and integrating the absorbed momentum in the direction of each face (Eq.~\ref{eqn:face.integrated.dp}). Regardless of the hydrodynamic grid, in moments-based methods (FLD/OTVET/M1) we must explicitly construct this ``effective grid'' on a sub-cell scale in order to obtain correct solutions (see \S~\ref{sec:moments.additional.problems}). And in hydrodynamic methods where the grid is anything but regular (e.g.\ AMR or moving-mesh or MFM/MFV methods), constructing faces and solving the relevant integrals for an arbitrary geometric distribution of mesh-generating points and source location is highly non-trivial. Moreover in SPH or finite-point methods (FPM) no explicit grid exists. We therefore present here a general method which can be used for {\em any} set of mesh-generating points (or ``particles'' in SPH/FPM), to construct a set of effective faces and solve Eq.~\ref{eqn:face.integrated.dp} in the immediate vicinity of a radiation source.

\begin{enumerate}

\item Every timestep $\Delta t_{a}$, for each source $a$ (at position ${\bf x}_{a}$), determine $L_{\nu}^{a}$ and $\Delta E_{\nu}^{a} = L_{\nu}^{a}\,\Delta t_{a}$. 

\item Identify ``neighboring'' gas surrounding the source. In regular-mesh methods this is straightforward; in mesh-free methods one identifies all gas elements within a ``search radius'' $H_{a}$ centered on $a$ ($|{\bf x}_{ba}| \equiv |{\bf x}_{b}-{\bf x}_{a}| < H_{a}$), and all elements for which $a$ is within their search radius ($|{\bf x}_{ba}| < H_{b}$).\footnote{The most common approach to define $H_{a}$ is via a target ``neighbor number'' $N_{\rm ngb} = (4\pi/3)\,H_{a}^{3}\,\bar{n}_{a}(H_{a})$, where $\bar{n}_{a}=\sum W({\bf x}_{ba},\,H_{a})$ and $W$ is an appropriate kernel function which integrates (over volume) to unity.}

\item Construct ``effective faces'' ${\bf A}_{ab}$ around the source. There are many possible choices for this, corresponding to different hydrodynamic methods. For example, a Voronoi tesselation (using $a$ and all $b$ as mesh-generating points). For the kernel-volume decomposition used in MFM/MFV methods \citep{hopkins:gizmo}: ${\bf A}_{ba} \equiv  \bar{n}_{a}^{-1}\,\bar{\bf q}_{b}({\bf x}_{a}) + \bar{n}_{b}^{-1}\,\bar{\bf q}_{a}({\bf x}_{b})$ where $\bar{\bf q}_{b}({\bf x}_{a}) \equiv {\bf E}_{a}^{-1} \cdot {\bf x}_{ba}\, W({\bf x}_{ba},\,H_{a})$ and ${\bf E}_{a} \equiv \sum_{c}\,({\bf x}_{ca} \otimes {\bf x}_{ca}) \,W({\bf x}_{ca},\,H_{a})$. In SPH, ${\bf A}_{ba} = [\bar{n}_{a}^{-2}\,\partial W(|{\bf x}|_{ba},\,H_{a})/\partial |{\bf x}|_{ba} + \bar{n}_{b}^{-2}\partial W(|{\bf x}|_{ba},\,H_{b})/\partial |{\bf x}|_{ba} ]\ \hat{\bf x}_{ba}$.

\item Solve the integral in Eq.~\ref{eqn:face.integrated.dp}, towards each face ${\bf A}_{ab}$. Assume $\rho$ and $\kappa_{\nu}$ are constant within each (sub-cell-scale) subdomain, and take ${\bf x}_{a}$ to be the coordinate origin. Then for single-scattering, ${\bf F}_{\nu}^{a}(r) = [L_{\nu}/(4\pi\,r^{2})]\,\exp{(-r/\mfp)}$ along each ray from the source (and we can add all sources independently). The integral can then be expressed as a set of vector weights $\bar{\bf w}_{ba}$:
\begin{align}
(\dot{\bf p}_{\nu})_{ab} &\approx f_{\rm abs}\,\frac{L_{\nu}^{a}}{c}\,\bar{\bf w}_{ba} \\ 
\label{eqn:vector.weight.normalized} \bar{\bf w}_{ba} &\equiv \frac{{\bf w}_{ba}}{\sum_{c}\,|{\bf w}_{ca}|} \\ 
\label{eqn:vector.weight.normalized.sub1} {\bf w}_{ba} &\equiv \omega_{ba}\, \sum_{+,\,-}\,\sum_{\alpha}\,(\hat{\bf x}_{ba}^{\pm})^{\alpha}\,\left( f_{\pm}^{\alpha} \right)_{a} \\ 
\label{eqn:vectornorm} \left( f_{\pm}^{\alpha} \right)_{a} &\equiv \left\{ \frac{1}{2}\,\left[1 +  \left( \frac{\sum_{c}\,\omega_{ca}\,|\hat{\bf x}_{ca}^{\mp}|^{\alpha}}{\sum_{c}\,\omega_{ca}\,|\hat{{\bf x}}_{ca}^{\pm}|^{\alpha}} \right)^{2}\right]\right\}^{1/2} \\
\label{eqn:solidangle}\omega_{ba} & \equiv \frac{1}{2}\,\left(1-\frac{1}{\sqrt{1+({\bf A}_{ba}\cdot \hat{\bf x}_{ba})/(\pi\,|{\bf x}_{ba}|^{2})}}\right) \approx \frac{\Delta\Omega_{ba}}{4\pi}
\end{align}
where the $\hat{\bf x}_{ca}^{\pm}$ are the positive or negative (singly-signed) projection vectors: 
\begin{align}
\label{eqn:vector.weight.def} \hat{\bf x}_{ba} &\equiv \frac{{\bf x}_{ba}}{|{\bf x}_{ba}|} = \sum_{+,\,-}\,\hat{\bf x}_{ba}^{\pm} \\ 
\label{eqn:vector.weight.def.sub1} (\hat{\bf x}^{+}_{ba})^{\alpha} &\equiv {|{\bf x}_{ba}|^{-1}}\,{\rm MAX}({\bf x}_{ba}^{\alpha},\,0)\,{\Bigr|}_{\alpha=x,\,y,\,z}\\
\label{eqn:vector.weight.def.sub2} (\hat{\bf x}^{-}_{ba})^{\alpha} &\equiv {|{\bf x}_{ba}|^{-1}}\,{\rm MIN}({\bf x}_{ba}^{\alpha},\,0)\,{\Bigr|}_{\alpha=x,\,y,\,z}
\end{align} 
and, for single-scattering, $f_{\rm abs} \approx 1 - \exp{(-|{\bf x}_{ba}|/\mfp)}$. For multiple-scattering from a single source, the expression above is identical up to $f_{\rm abs}$ -- if we assume a gray opacity and perfect re-radiation (within the single cell sub-domain) then $f_{\rm abs} \approx |{\bf x}_{ba}|/\mfp$. 

These expressions are derived in \citet{hopkins:sne.methods} (up to the trivial addition here of calculating $f_{\rm abs}$); while non-trivial they have three key properties. (1) They maintain manifest conservation of linear momentum. (2) They give fluxes which are statistically isotropic in the rest frame of the source (i.e.\ the coupled momenta are not numerically biased in any particular direction, regardless of the position of the source within the cell or the position/distribution of mesh-generating points/particles). (3) They approximate, as closely as possible without a computational expensive exact numerical quadrature, the exact integral of Eq.~\ref{eqn:face.integrated.dp}. 

\item Verify conservation: one should ensure total momentum and energy are conserved, regardless of how Eq.~\ref{eqn:face.integrated.dp} is solved. For an isotropic source in a uniform density field, this means ensuring the {\em total linear} momentum coupled vanishes (the momentum-coupling is symmetric). If one solves Eq.~\ref{eqn:face.integrated.dp} {\em exactly} at all points, {\em and} has an exact solution for ${\bf F}_{\nu}$ at all points, {\em and} has a set of faces which form an exactly-closed convex hull, then this is guaranteed mathematically, but numerically these conditions are not usually all satisfied. Here, this step is included above: the $f_{\pm}^{\alpha}$ terms in Eq.~\ref{eqn:vectornorm} are numerically evaluated in a first-loop over the neighbors, to calculate ${\bf w}_{ba}$, and represent the ``correction'' (vector re-normalization) needed to guarantee conservation (one can verify that if the conditions above were met, then one would always have $f_{\pm}^{\alpha}=1$ exactly). 

\item Couple the momentum to the gas fully-conservatively, either as direct fluxes/forces into each neighbor $b$, or in a Riemann problem (just adding the flux to the ``other'' fluxes evaluated at the face in that step).

\end{enumerate}

The procedure above resolves the primary issues in the text, on arbitrary grid, for arbitrary source positions. Note that we have calculated the face-integrated coupling in the vicinity of the source. Far away from the source, one can extend a method like this if desired. However, far from a source, the direction of the flux is not strongly varying within a single cell -- so simply using the cell-centered average flux (equivalently, assuming the net flux ${\bf F}_{\nu}({\bf x})$ at all positions ${\bf x}$ inside a cell $b$, far from any sources, points in approximately the same direction) is not a significant source of error in this regime.

\end{appendix}

\end{document}